\documentclass[11pt]{article}
\usepackage{moriond}
\usepackage{graphicx}

\def\Journal#1#2#3#4{{#1} {\bf #2}, #3 (#4)}

\def\PLB{{\em Phys. Lett.}  B}
\def\PRL{\em Phys. Rev. Lett.}
\def\PRD{{\em Phys. Rev.} D}

\def\be{\begin{equation}}
\def\ee{\end{equation}}
\def\bea{\begin{eqnarray}}
\def\eea{\end{eqnarray}}

\begin{document}
\vspace*{4cm}
\title{STANDARD MODEL FITS}

\author{ S. ROTH }

\address{III.~Physikalisches~Institut, RWTH Aachen\\
         D-52056 Aachen, Germany}

\maketitle

\abstracts{
The status of the electroweak precision measurements as of winter 2004 
and the global test of the Standard Model are discussed.
Important input data are the precision variables measured on
the Z resonance at LEP and SLC and the measurements of the W mass
at LEP~2 and Tevatron.
A new combination of Tevatron experiments CDF and D0 on the top mass 
allows to set constraints on the radiative corrections and therefore to
put improved limits on the mass of the Higgs boson.
Additionally the impact of the NuTeV result on the weak mixing angle 
and the status of the calculation of the hadronic vacuum polarization
$\Delta \alpha_{\rm had}$ are discussed.
}

\section{Measurements at the Z Resonance}

\begin{figure}
\includegraphics[height=0.3\textheight]{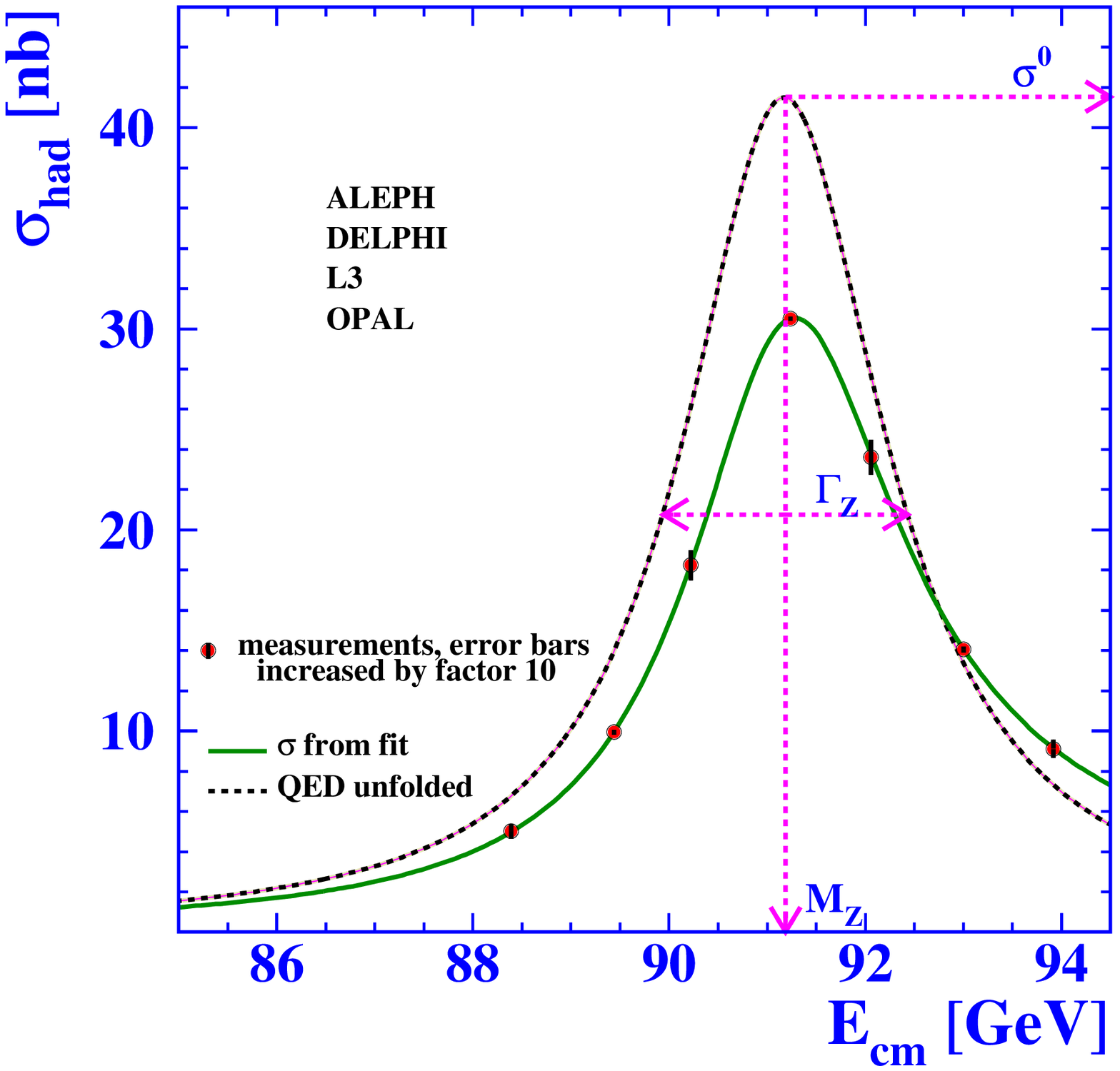}
\hfill
\includegraphics[height=0.3\textheight]{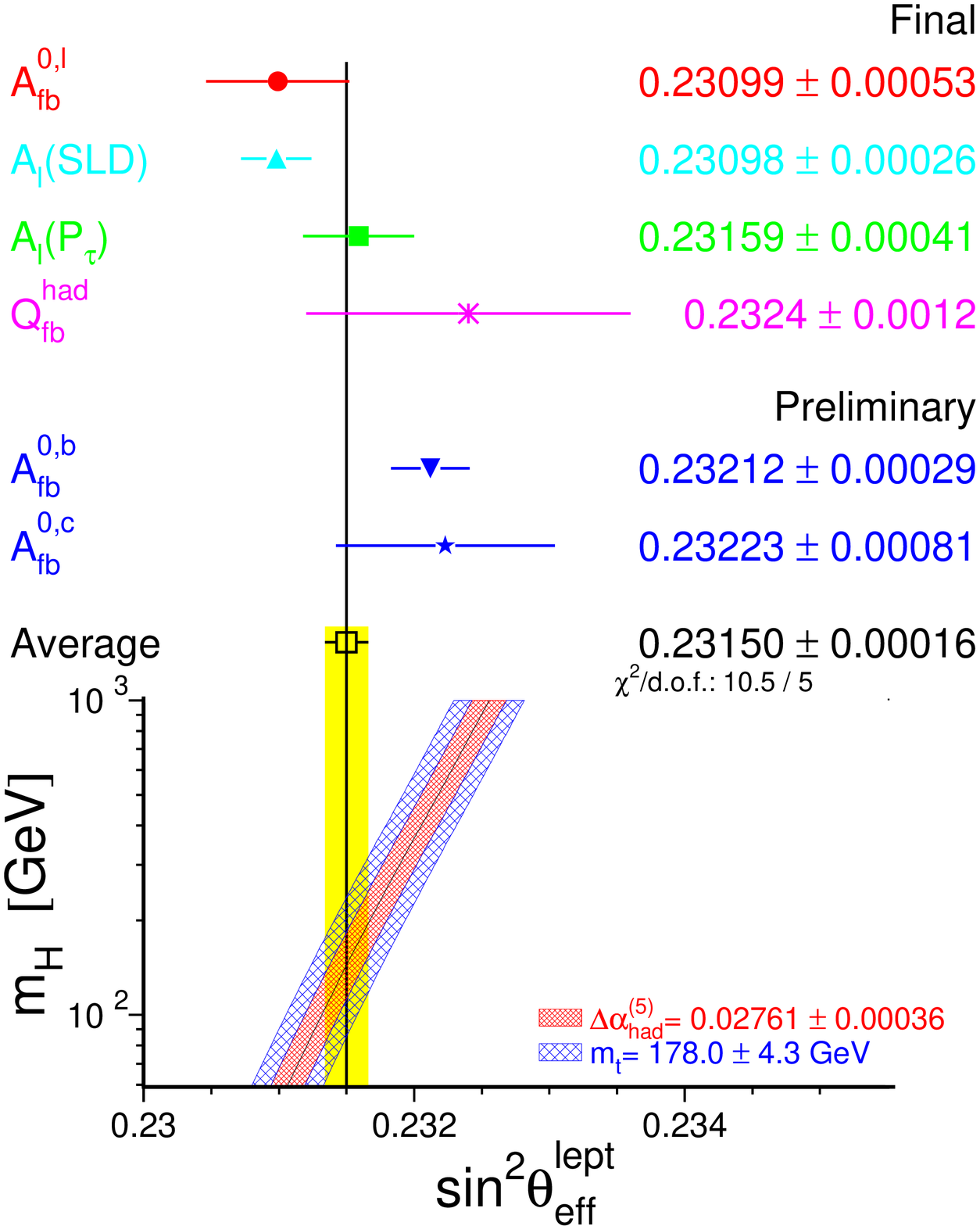}
\caption{Measurement of the total hadronic cross cross section on the Z resonance (left);
measurements of the weak mixing angle from Z decays and comparison with the SM prediction (right).}
\label{fig:mz-sef2}
\end{figure}

The Standard Model (SM) is confirmed at the permille level using 
electroweak precision data.
The analysis of electron-positron collisions at centre-of-mass energies
around the Z resonance has delivered a wealth of precisely measured 
electroweak observables.
One observable is the mass of the Z, which is derived mainly from
the measurement of the total hadronic cross section as shown in 
Figure~\ref{fig:mz-sef2} (left).
Today the Z mass is known with a precision of 23~ppm~\cite{lep-z}:
\[
   m_{\rm Z} = 91.1875 \pm 0.0021 \; {\rm GeV} \; .
\]

Other observables are related to the coupling of the weak current
to the fermions and can therefore be expressed in terms of the 
electroweak mixing angle.
The different values of $\sin^2 \theta^{\rm lept}_{\rm eff}$ derived 
from the measurements of the lepton forward-backward asymmetry, $A^{\rm 0,l}_{\rm fb}$,
the left-right asymmetry at SLD, $A_{\rm l}({\rm SLD})$, the $\tau$ polarization
at LEP, $A_{\rm l}(P_\tau)$, and the forward-backward asymmetry of the
b-quark and c-quark final state, $A^{\rm 0,b}_{\rm fb}$ and $A^{\rm 0,c}_{\rm fb}$
are compared in Figure~\ref{fig:mz-sef2} (right).
A difference of 2.9~standard deviations is observed between the two most
precise measurements, the left-right asymmetry and
the b-quark forward-backward asymmetry.
Combining all measurements results in an accuracy at the sub permille 
level~\cite{winter04}: 
\[
   \sin^2 \theta^{\rm lept}_{\rm eff} = 0.23150 \pm 0.00016 \; .
\]
Comparing measurement and theory prediction shows a
preference for a light Higgs mass.

\section{Measurement of the W mass}

Within the SM the Z mass, $m_{\rm Z}$, the W mass, $m_{\rm W}$, 
and the weak mixing angle, $\sin^2 \theta_{\rm w}$, are 
related by
\[
   \cos \theta_{\rm w} = \frac{m_{\rm W}}{m_{\rm Z}},
\]
which is naturally predicted by the Higgs mechanism.
To test this relation a precise measurement
of the W mass is mandatory.

Since 1996 the W mass is measured at LEP~2 studying the
four-fermion production through
${\rm e}^+{\rm e^-} \to {\rm W}^+{\rm W}^- \to f \bar{f} f \bar{f}$.
The W bosons are reconstructed from the 
measured momenta of the observed final state fermions.
To extract the W mass the  invariant mass spectrum is compared 
to the one obtained from Monte-Carlo events with complete 
detector simulation.
In the four-jet final state momentum exchange between the
decay products of two W bosons may occur due to colour reconnection
or Bose-Einstein effects in the non-perturbative phase of
the jet formation.
Because of these additional uncertainties compared to the
semileptonic channel the weight of the four jet channel in the 
LEP average is currently only about 10\%.
Further details on the measurement of the W mass at LEP
are given elsewhere~\cite{lep-wmass}.

At hadron colliders leptonic W decays with electrons and muons
are selected and the transverse mass is calculated.
The transverse mass, i.e. the invariant mass of the transverse 
momentum of the charged lepton and the missing momentum vector 
in the plane transverse to the beam, is not affected by the
unknown missing momentum along the beam axis.
Recently the experiments CDF and D0 performed a 
precise measurement of the W mass using the Run~I data set
of the Tevatron collider \cite{tev-wmass}.
The precision of the Tevatron W mass measurement is currently
limited by data statistics.
The uncertainty in the lepton energy scale gives the largest 
contribution to the systematic error.

\begin{figure}
\includegraphics[height=0.25\textheight]{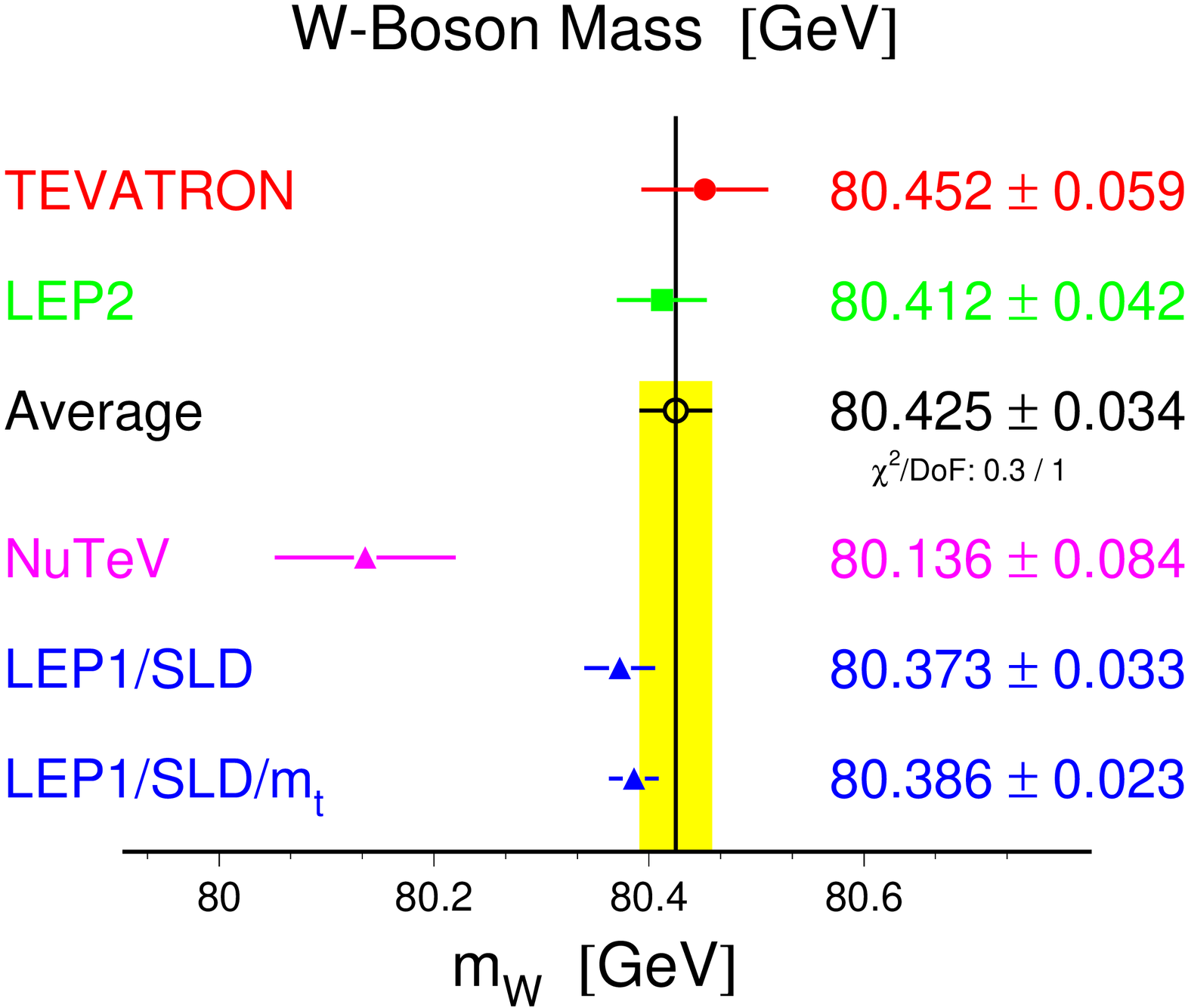}
\hfill
\includegraphics[height=0.25\textheight]{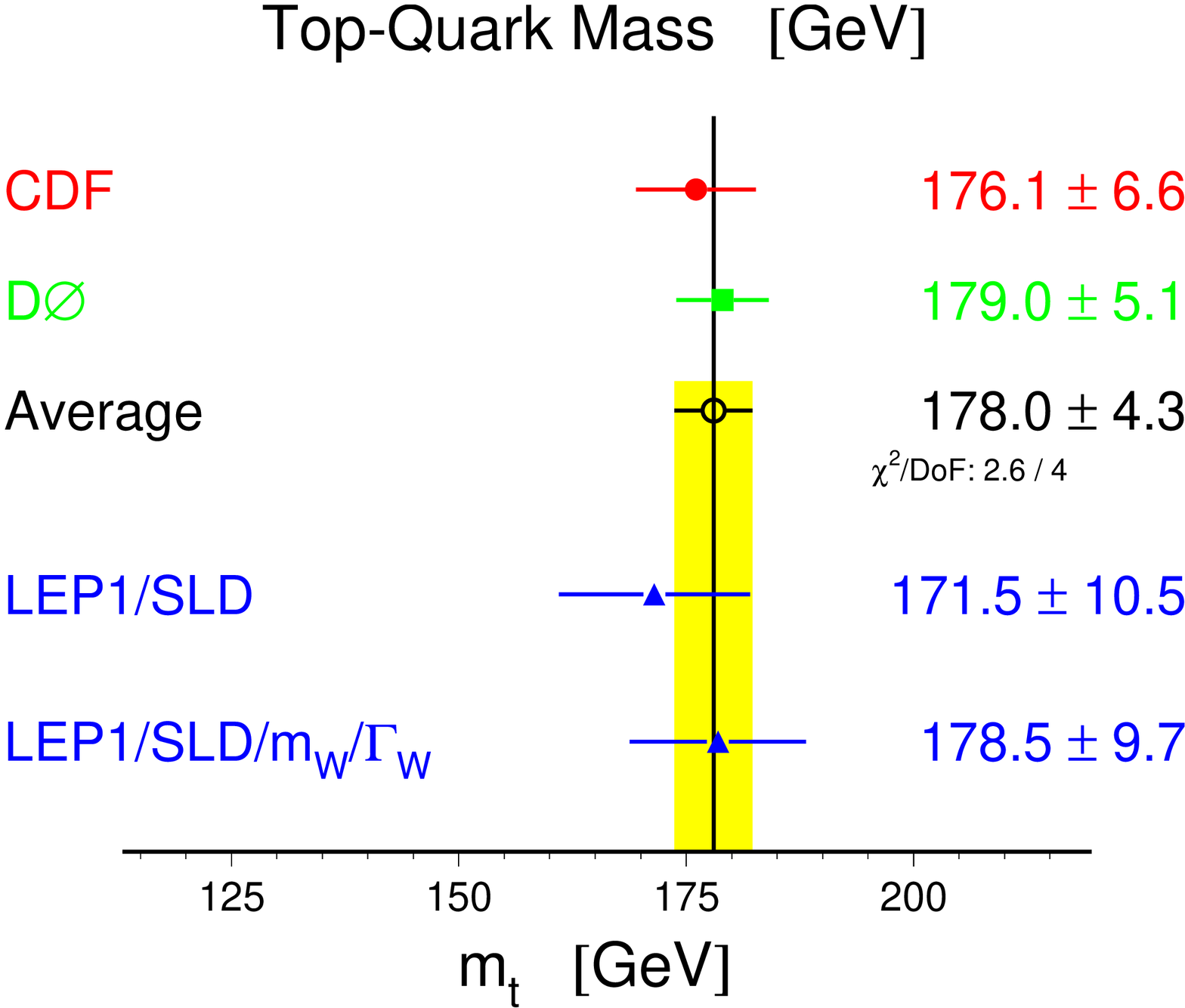}
\caption{Measurement of the W mass at LEP and Tevatron (left);
measurement of the Top mass at the Tevatron and comparison
with the prediction from LEP data(right).}
\label{fig:mw-mt}
\end{figure}

The results of the Tevatron and LEP experiments on the W mass
are in good agreement as shown in Figure~\ref{fig:mw-mt} (left).
All direct W mass measurements, 
of which most of them are still preliminary, 
result in a world average of:
\[
   80.425 \pm 0.034 \; {\rm GeV} \; .
\]

Another less precise indirect measurement of the W mass
is coming from the measurement of neutrino nucleon scattering.
Measurements of the NuTeV collaboration~\cite{nutev} show a 
deviation from the world average of about three standard deviations.
But recent theoretical studies~\cite{nutev-theo} suspect that 
the uncertainties due to QCD corrections and due to 
electroweak radiative correction may be underestimated 
in this analysis.

\section{Other Observables}
\label{other}

Electroweak radiative corrections have been calculated up to the two-loop level,
but there accuracy is limited by the experimental uncertainties
for the masses of the top quark and the unknown mass of the Higgs boson.
A test of the quantum structure of the SM therefore requires
a precise knowledge of the top quark mass, as the radiative corrections
depent quadratically on this parameter.

In the year 1995 the experiments at the Tevatron collider discovered
in the mass range predicted by the electroweak measurements at LEP.
They observe the top quark in the reaction 
${\rm p} \bar{\rm p} \to {\rm t} \bar{\rm t} X 
\to {\rm b} \bar{\rm b} {\rm W}^+ {\rm W}^- X$
If the W boson decays into two quarks, the mass of the top quark can
be reconstructed from the invariant mass of the b jet and the two
jets coming from the W decay.
The results based on data collected in Run I and partly in Run II
have recently been combined~\cite{tevewwg}:
\[
  m_{\rm t} = 178.0 \pm 2.7({\rm stat}) \pm 3.3({\rm syst}) \; {\rm GeV} \; .
\]
The new and improved preliminary Run-I based result of the D0 collaboration
in the lepton-plus-jets channel is included in this value. 
A comparison between the different measurements of the top mass
is shown in Figure~\ref{fig:mw-mt} (right).

The calculation of the electromagnetic coupling constant $\alpha(m_{\rm Z})$
at the energy scale of the Z mass is dominated by the knowledge on the
hadronic vacuum polarisation $\Delta \alpha_{\rm had}$.
This part can not be calculated by pertubation theory, but has to be
extracted using the total hadronic cross section in ${\rm e}^+ {\rm e}^-$
collisions.
A very important part is coming from the region of the $\rho$ resonance.
Precise measurements of the CMD-2 detector get now confirmed by new
results from KLOE.
The most recent compilation~\cite{alpha} gives
\[
   \Delta \alpha^{(5)}_{\rm had}({m_{\rm Z}}) = 0.02761 \pm  0.00036 \; .
\]

\section{Interpretation within the Standard Model and Higgs mass analysis}

The details of the combination of the electroweak precision data and
the global SM fit is described elsewhere~\cite{lepewwg}.
The fit results in a rather poor $\chi^2$, but excluding the low $Q^2$
experiments gives a much better fit quality with a 27\% probability.

As stated above, the electroweak radiative corrections
include a term proportional to the logarithm of the Higgs mass.
Assuming the validity of the SM one can try to extract this term
from a global fit of all electroweak precision measurement.
This allows to predict the mass of the Higgs boson.

\begin{figure}
\includegraphics[height=0.35\textheight]{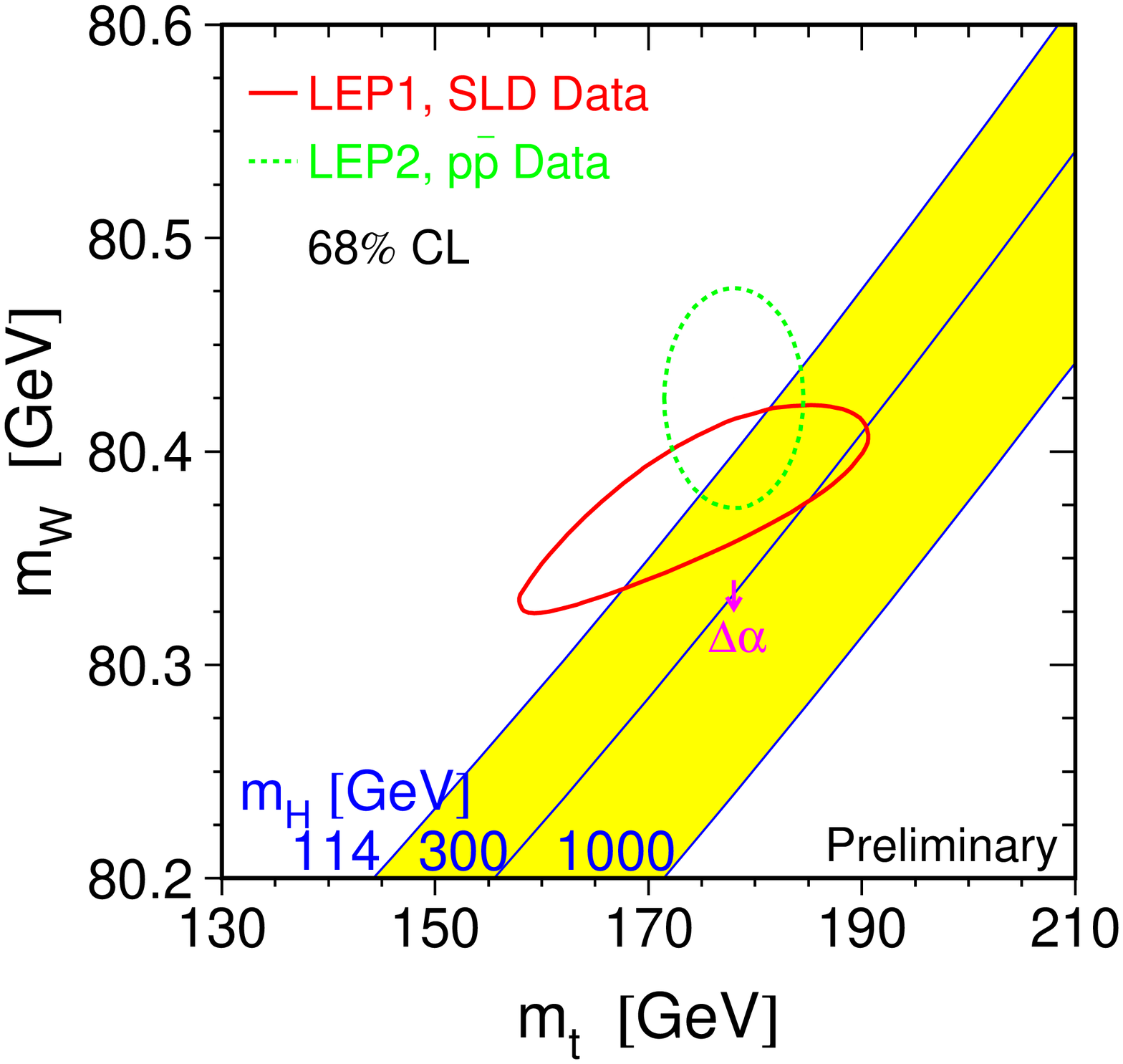}
\hfill
\includegraphics[height=0.35\textheight]{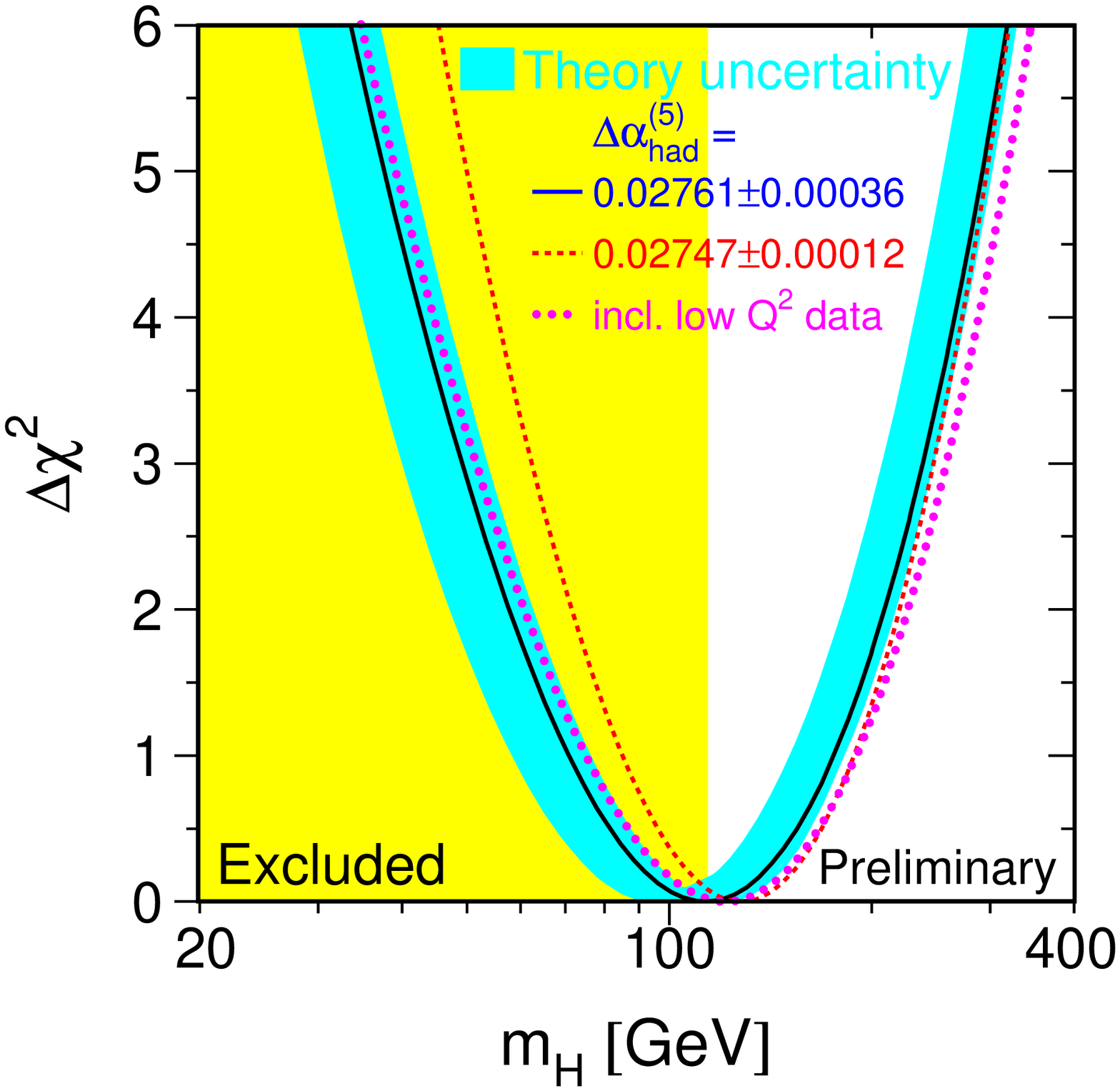}
\caption{Comparison of direct mass measurements with prediction
from the SM and from electroweak precision data (left);
Prediction of the Higgs mass from global SM fit (right).}
\label{fig:sm-fit}
\end{figure}

In Figure~\ref{fig:sm-fit} the constraint from the electroweak precision
measurements performed at LEP~1 and SLD in the $m_{\rm W}$ vs. $m_{\rm t}$
plane is shown together with the direct measurement of the W and the top mass.
Additionally the SM prediction using $G_{\rm F}$ from muon decay is plotted
for different Higgs masses.
Both the indirect and the direct measurements prefer a low Higgs mass.
The fit of SM modell prediction to all electroweak data with
the Higgs mass as the only free parameter results in a $\chi^2$ curve
as shown in Figure~\ref{fig:sm-fit}.
It gives a central value of 117~GeV for the Higgs mass, which
is consistent with the direct searches for the Higgs excluding masses
below 114.4~GeV~\cite{higgs}.
At 95\% C.L.\ an upper bound on the Higgs mass of 251~GeV is set.

\end{document}